\documentclass[sigconf]{acmart} 

\AtBeginDocument{%
  \providecommand\BibTeX{{%
    \normalfont B\kern-0.5em{\scshape i\kern-0.25em b}\kern-0.8em\TeX}}}

\copyrightyear{2022}
\acmYear{2022}
\setcopyright{acmlicensed}\acmConference[JCDL '22]{The ACM/IEEE Joint Conference on Digital Libraries in 2022}{June 20--24, 2022}{Cologne, Germany}
\acmBooktitle{The ACM/IEEE Joint Conference on Digital Libraries in 2022 (JCDL '22), June 20--24, 2022, Cologne, Germany}
\acmPrice{15.00}
\acmDOI{10.1145/3529372.3530953}
\acmISBN{978-1-4503-9345-4/22/06}

\usepackage{subfigure}
\usepackage{caption}
\usepackage{tabularx}
\usepackage{booktabs}
\usepackage{multirow}
\usepackage{multicol}

\begin{document}


\title{How Does Author Affiliation Affect Preprint Citation Count? Analyzing Citation Bias at the Institution and Country Level} 



\author{Chifumi Nishioka}
\affiliation{%
  \institution{National Institute of Informatics}
  \city{Tokyo}
  \postcode{101-8430}
  \country{Japan}
}
\email{cnishioka@nii.ac.jp}

\author{Michael Färber}
\orcid{0000-0001-5458-8645}
\affiliation{%
  \institution{Karlsruhe Institute of Technology}
  \city{Karlsruhe}
  \postcode{76133}
    \country{Germany}
}
\email{michael.faerber@kit.edu}

\author{Tarek Saier}
\orcid{0000-0001-5028-0109}
\affiliation{%
  \institution{Karlsruhe Institute of Technology}
  \city{Karlsruhe}
  \postcode{76133}
    \country{Germany}
}
\email{tarek.saier@kit.edu}

\renewcommand{\shortauthors}{Nishioka et al.}

\begin{abstract}
Citing is an important aspect of scientific discourse and important for quantifying the scientific impact quantification of researchers. Previous works observed that citations are made not only based on the pure scholarly contributions but also based on non-scholarly attributes, such as the affiliation or gender of authors. In this way, citation bias is produced. Existing works, however, have not analyzed preprints with respect to citation bias, although they play an increasingly important role in modern scholarly communication. 
In this paper, we investigate whether preprints are affected by citation bias with respect to the author affiliation. 
We measure citation bias for bioRxiv preprints and their publisher versions at the institution level and country level, using the Lorenz curve and Gini coefficient. 
This allows us to mitigate the effects of confounding factors and see whether or not citation biases related to author affiliation have an increased effect on preprint citations. 
We observe consistent higher Gini coefficients for preprints than those for publisher versions. Thus, we can confirm that citation bias exists and that it is more severe in case of preprints.  
As preprints are on the rise, affiliation-based citation bias is, thus, an important topic not only for authors (e.g., when deciding what to cite), but also to people and institutions that use citations for scientific impact quantification (e.g., funding agencies deciding about funding based on citation counts). 

\end{abstract}


\keywords{preprint, bibliometrics, citation, bias}


\maketitle

\section{Introduction}
\label{sec:introduction}

Citing is an important aspect of scientific discourse and important for quantifying the scientific impact quantification of researchers. Widely used importance metrics, such as the citation count and the h-index~\cite{hirsch2005index}, are based on citations. 
They are sometimes used to judge the quality of research presented by an article~\cite{nieminen2006relationship}. 
However, several works have observed that publications are cited not only based on the pure scholarly contributions but also based on non-scholarly attributes such as gender, author affiliation, and funding. 
For instance, articles authored by women might be under-cited~\cite{caplar2017quantitative,dion2018gendered,teich2021citation}. 
Such distortions concerning citations---also called ``citation bias''---can distort the perception of available scholarly contributions among users of publications~\cite{jannot2013citation}. 

While citation bias has been studied in regular journal articles~\cite{dion2018gendered,teich2021citation,VanNoorden2019HundredsDatabase,Aksnes2003ASelf-citation}, citation bias in preprints---completed scientific manuscripts that are uploaded by the authors to a public server without formal review~\cite{berg2016preprints}---has not been investigated. 
However, preprints play an increasingly important role in modern scholarly communication. 
Several preprint servers have emerged within the last decades~\cite{xie2021preprint}, covering various disciplines: arXiv in physics, mathematics, and computer science, bioRxiv in biology, medRxiv in medicine, and SSRN in social science. 
Various works have observed benefits of preprints, such as early disclosure, wider dissemination~\cite{sarabipour2019value} resulting in a higher number of citations~\cite{davis2007does,fraser2020relationship,feldman2018citation}, and creating opportunities for collaborations~\cite{kleinert2018preprints,sarabipour2019value,penfold2020technical}. 
In the recent COVID-19 pandemic, preprints have received even greater scientific and public engagement~\cite{fraser2021evolving}. 

In this paper, we investigate if preprints are affected by citation bias concerning the author affiliation.
We focus on the author affiliation, as a survey by Soderberg et al.~\cite{soderberg2020credibility} observed that 35\% of respondents consider the author's institution as extremely or very important to assess the credibility of preprints. 
Therefore, we assume that author affiliation has an influence on the citation counts of preprints. 
We verify the existence of citation bias by computing citation inequality.
To this end, we measure to which degree the number of citations that preprints and their publisher versions receive is unequally distributed. 
Specifically, we measure citation bias with regard to author affiliation on the institution level and country level. 
Comparing differences in the citation inequality between preprints and their respective publisher versions allows us to mitigate the effects of confounding factors and see whether or not citation biases related to author affiliation have an increased effect on preprint citations.
Conclusions drawn from this type of investigation are based on the assumption that the process of peer-review and formal publication is generally perceived as an assurance of quality \cite{nieminen2006relationship} and therefore ``levels the playing field'' among articles in terms of citability.

We examine citation bias in bioRxiv, a preprint server in the field of biology, because preprints deposited to bioRxiv provide sufficient information regarding author affiliations. 
We analyze citations of more than 36,000 preprints deposited between November 2013 (i.e., the launch of bioRxiv) and June 2019 and their publisher versions. 
We use the COCI (OpenCitations Index of Crossref open DOI-to-DOI references)~\cite{heibi2019software} as citation data. 
To measure citation inequality, we calculate the Gini coefficients $G$, following previous studies~\cite{nielsen2021global,ettarh2021analysis}. 
In our analysis, we can confirm a citation bias, especially for preprints at different affiliation levels (i.e., institutions, countries), as we find that preprints have twice the citation inequality as the publisher versions (e.g., $G=0.23$ for preprints and $G=0.12$ for publisher versions at the institution level). 
Furthermore, we observe larger citation inequalities for preprints than those for publisher versions in different journal types that are mega-journals\footnote{Mega-journals are journals that solely focus on scientific trustworthiness~\cite{bjork2018evolution} in the process of peer-review, compared to other journals.}, disciplinary journals, and prestigious journals (e.g., Nature and Science). 

Preprints begin to be increasingly considered in various contexts, such as funding applications and recruitment. Therefore, citations of preprints gain in importance. 
Given our results, funding agencies or referees are advised to be even more careful when they apply citation-based metrics to preprints for their assessment and judgment than applying to journal articles. 

The reminder of the paper is organized as follows: In the subsequent section, we describe the related works. Thereafter, we show the procedure of the data collection for the analysis in Section~\ref{sec:data-collection} and the analysis methods in Section~\ref{sec:evaluation}. Section~\ref{sec:results} presents the results of the analysis. In Section~\ref{sec:discussion}, we discuss and outline the analysis, its limitation, and future direction, before concluding the paper in Section~\ref{sec:conclusion}. 

We provide our dataset and source codes used for the analysis online~\cite{src}.

\section{Related Work}
\label{sec:related-word}

In this section, we outline related works. We first describe studies related to preprint characteristics. Thereafter, we mention related analyses that investigate factors affecting citations and citation bias. Finally, we show different studies in terms of biases in academia. 

\subsection{Preprint Characteristics} 

The advent of preprint servers has brought about a change in citation behavior. For instance, researchers who no longer need the publication of papers from publishers for their career may skip the review process and no longer publish in journals of publishers \cite{DBLP:journals/jasis/KimPWS20}. 

Soderberg et al.~\cite{soderberg2020credibility} conducted a survey of 
almost 4,000 
researchers across different disciplines to determine the importance of different cues for assessing the credibility of preprints and preprint services. 
They found that cues related to information about open science content (e.g., links to available material, data, and scripts) as well as independent verification of author claims (e.g., information about independent reproductions) were rated as highly important for judging preprint credibility. In comparison, peer views and author information were rated as less important. 
35\% of respondents marked the author's institution as extremely or very important, and 28\% of respondents answered moderately important.  
This motivated us to consider citation bias with respect to the author affiliations.


\subsection{Factors Affecting Citations and Citation Bias} 

Tahamtan et al.~\cite{tahamtan2016factors} outlined factors that affect the number of citations. Specifically, they identified three categories with 28 factors to be related to the number of citations: paper related factors (e\,.g., quality of paper, document type), journal related factors (e\,.g., journal impact factor), and author(s) related factors (e\,.g., number of authors). They concluded that some factors, such as the journal impact factor, international cooperation, and number of authors, are more strongly correlated with the number of citations than the other factors. 
These factors include non-scholarly attributes such as gender, author affiliation, and funding. 
This phenomenon of inequality can be referred to as ``citation bias,'' and it distorts the perception of available scholarly contributions among users of articles~\cite{jannot2013citation}. 
Several studies found that papers authored by women might be under-cited~\cite{caplar2017quantitative,dion2018gendered,teich2021citation}, while Copenheaver et al.~ \cite{copenheaver2010lack} did not observe citation bias based on gender. 
Lou and He~\cite{lou2015does} observed a significant negative correlation between the reputation of author affiliations (i\,.e., rank of an affiliation at the U.S. News Best Global University Subject rankings) and uncitedness of journal articles. 



\subsection{Biases in Academia}

Citation bias has been analyzed in two main contexts in the literature: to explain the scholars' self-citation behavior \citep{VanNoorden2019HundredsDatabase,Aksnes2003ASelf-citation}, and 
to show that scholars cite papers but disproportionally criticize papers or specific claims less often. 
%
%
Besides the citation bias, also other kinds of biases in academia have been studied. 
For instance, Liang et al. \cite{Liang2016} discussed the recommender systems' ``exposure problem,'' which can result in frequent recommendation of popular scientific articles. 
Salman et al.~\cite{Salman2020IncorporatingRecommendation} observed gender and racial biases and location-based biases in academic expert recommendations, used to find reviewers or to assemble a conference program committee. 
Polonioli et al.~\cite{Polonioli2020TheSystems} claimed 
that recommender systems might put users in information bubbles by isolating them from exposure to different academic viewpoints, creating a self-reinforcing bias damaging to scientific progress. Finally, \citet{Gupta2021CorrectingRecommendation} found that scholarly recommender systems are biased as they underexpose users to equally relevant items. 
A paper that tackles the popularity bias of recommending scientific articles \citep{Wang2011CollaborativeArticles}  won the Test of Time award at the KDD 2021 conference.\footnote{\url{https://kdd.org/awards/view/2021-sigkdd-test-of-time-award-winners}, last accessed on 2021-12-14} 
All this shows that bias is an important and timely topic to consider.

\section{Data Collection}
\label{sec:data-collection}

This section describes how we collect preprints, their respective publisher versions, and citation data. 

\subsection{Preprints and their Publisher Versions}
\label{sec:preprints-and-their-publisher-versions}

\paragraph{Preprints}
bioRxiv is a widely used preprint server in biology and provides---in contrast to other preprint servers, such as arXiv---easy access to the author affiliation information of the preprints.\footnote{In arXiv, we need to extract author affiliations from the body of PDF or LaTex. On the other hand, bioRxiv provides metadata and text in a unified JATS format, which makes easy to retrieve author affiliations.}
We therefore harvest metadata of all bioRxiv preprints submitted between November 2013 (i.e., the launch of bioRxiv) and June 2019 via the bioRxiv API.\footnote{\url{https://api.biorxiv.org/}, last accessed on 2022-04-29} 

We harvest metadata until June 2019 to ensure that all preprints and their respective publisher versions have at least a 24-month period to receive citations after publication of preprints. 
Therefore, this paper does not cover preprints and publisher versions that are related to COVID-19. 

In total, we retrieve 73,946 records. 
After removing duplicate records we obtain 73,920 records. 


%

Thereafter, using a bioRxiv metadata field ``JATS URL'' in the records, we download JATS~\cite{Huh2014} XML files for each of the 73,920 records.  
On bioRxiv, authors can update their preprints. Therefore, some submissions are available in several versions.
In this paper, we only use the metadata and JATS XML files of the first version of a preprint, as we assume that metadata, such as the author information, do not change between versions. 
Following above steps, we acquire metadata and JATS XML files of 53,240 preprints.

\paragraph{Publisher versions}
We identify the publisher version of a preprint using a bioRxiv metadata field that provides a link to the publisher version. These links, DOIs in most cases, are identified and stored as bioRxiv metadata automatically whenever a corresponding author confirms the publication of a preprint via email.\footnote{\url{https://www.biorxiv.org/about/FAQ}, last accessed on 2021-12-14} 
We fetch the metadata of publisher versions, such as journal information\footnote{We use the fields \texttt{journal\_name} and \texttt{journal\_issn\_l}.} and publication month,\footnote{We use the field \texttt{published\_date} that corresponds to the field \texttt{created} in Crossref. } from Unpaywall, using the DOIs.
We filter out 50 publisher versions whose publication month cannot be identified and 471 publisher versions that have been published before preprint publication. 

Following above procedure, we identify 36,651 pairs of a preprint and its respective publisher version. 

\paragraph{Author Information}
We obtain author information including affiliations from the JATS XML files of preprints. 
We assume that the author information is identical in a preprint and its publisher version. 
We specifically fetch the following information: 
\begin{itemize}
    \item the order of authors
    \item whether an author is corresponding author
    \item each author's affiliation(s) 
\end{itemize}

There are 23 pairs where author information is unavailable. We exclude them from the analysis. Thus, we finally get 36,628 pairs of a preprint and its respective publisher version. 
For the 36,628 pairs, there are 260,231 authors. 

Author affiliations appear in different variations (e\,.g., ``MIT'' and ``Massachusetts Institute of Technology'') in JATS XML files. 
We normalize author affiliations using the Research Organization Registry (ROR)~\cite{lammey2020solutions}. 
The ROR is a community-led registry of identifiers for research organizations.\footnote{\url{https://ror.org/about/}, last accessed on 2021-12-14}  
It provides an API, which allows to retrieve, search, and filter the organizations indexed in the ROR.
We identify the corresponding ROR entity for each author affiliation string using this API.\footnote{\url{https://github.com/ror-community/ror-api}, last accessed on 2021-12-14}  
Although strings of some author affiliations are marked up with ``institution'' and ``country,''\footnote{An example: <institution>University of Minnesota</institution>, Minneapolis, MN 55455, <country>USA</country>} we used full strings of author affiliation as queries for consistency. 
The ROR API returns a list of organizations that are matched to a query sorted by confidence scores. 
We use the returned organization with the highest confidence score and if the field ``chosen'' (i.e., binary indicator of whether the score is high enough to consider the organization correctly matched) is true. 
For the 260,231 authors, there are 335,188 affiliations. Among them, 273,804 affiliations (81.69\%) can be linked to a ROR entity. 
In our analysis, we consider citation bias on the institution and country level. Thus, we use the name and country information of the organizations.

\subsection{Citation Data}

As citation data, we used the COCI (OpenCitations Index of Crossref open DOI-to-DOI references)~\cite{heibi2019software}. 
The citation data of the COCI are originally from publishers, thus they are of high quality. 
We used the COCI CSV dataset Version 11 released on 2021-09-03\footnote{\url{https://doi.org/10.6084/m9.figshare.6741422.v11}, last accessed on 2021-12-14} that lists pairs of DOIs denoting citations. 
The 36,628 preprints and their publisher versions receive 331,839 citations in total in the given 24 months. 

\section{Analysis Methods} 
\label{sec:evaluation}

In this section, we describe how we count the number of citations and how we identify citation bias. 

\subsection{Citations} 
Following Thelwall~\cite{thelwall2016discretised} and Fraser et al.~\cite{fraser2020relationship}, we log-transform the number of citations of an article (i\,.e., preprint, publisher version) after an addition of 1, to reduce the influence of articles with a high number of citations. 
Finally, we take the arithmetic mean of the log-transformed number of citations of all articles with respect to an affiliation (i.e., institution, country) as
\begin{equation}
	c_{m} = \frac{1}{n}\sum_{i=1}^{n}log(c_i + 1),
\end{equation}
where $n$ refers to the number of articles of an affiliation and $c_{i}$ is the number of citations of an article of an affiliation.

\begin{figure}[tb]
  \includegraphics[width=\columnwidth]{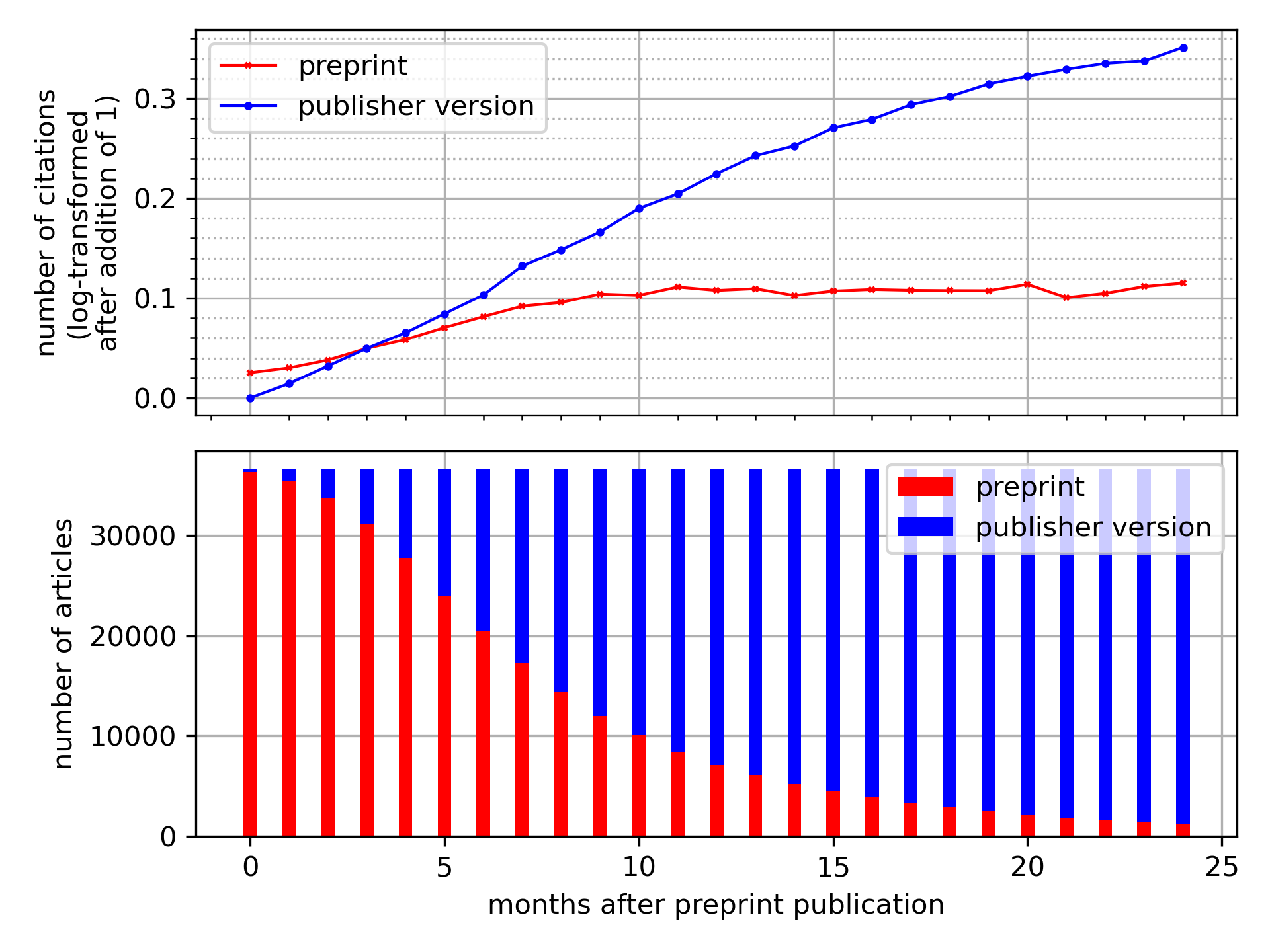}
  \caption{Number of citations (log-transformed after addition of 1) and articles considering preprints and publisher versions.}
  \label{fig:time_articles_citations}
\end{figure}

\subsection{Citation Bias} 

We examine citation bias concerning author affiliations at institution level and country level by measuring to which degree the number of citations that preprints and their publisher versions receive is unequally distributed. 
We assume that comparing differences in citation inequality between preprints and their publisher versions allows us to mitigate the effects of confounding factors and see whether or not citation biases related to author affiliation have an increased effect on preprint citations. 

Specifically, we plot the Lorenz curve and calculate the Gini coefficient~\cite{dorfman1979formula} to measure inequality in the number of citations, as used by authors of similar studies, such as Nielsen and Andersen~\cite{nielsen2021global}. 
In this paper, the Lorenz curve presents the distribution of citations accumulated across different affiliations, where it shows for the bottom $x$\% of affiliations, what percentage ($y$\%) of the total number of citations they received. If the distribution of the number of citations for different affiliations is perfectly equal, the Lorenz curve is depicted by the straight line $y = x$. 
The Gini coefficient is calculated as the ratio of the area between the line of perfect equality and the observed Lorenz curve to the area between the line of perfect equality and the line of perfect inequality. 
The Gini coefficient can range from 0 to 1. A higher Gini coefficient indicates a high degree of inequality in the distribution. 

\begin{figure}
	\subfigure[institution, first author]{%
		\includegraphics[clip, width=0.49\columnwidth]{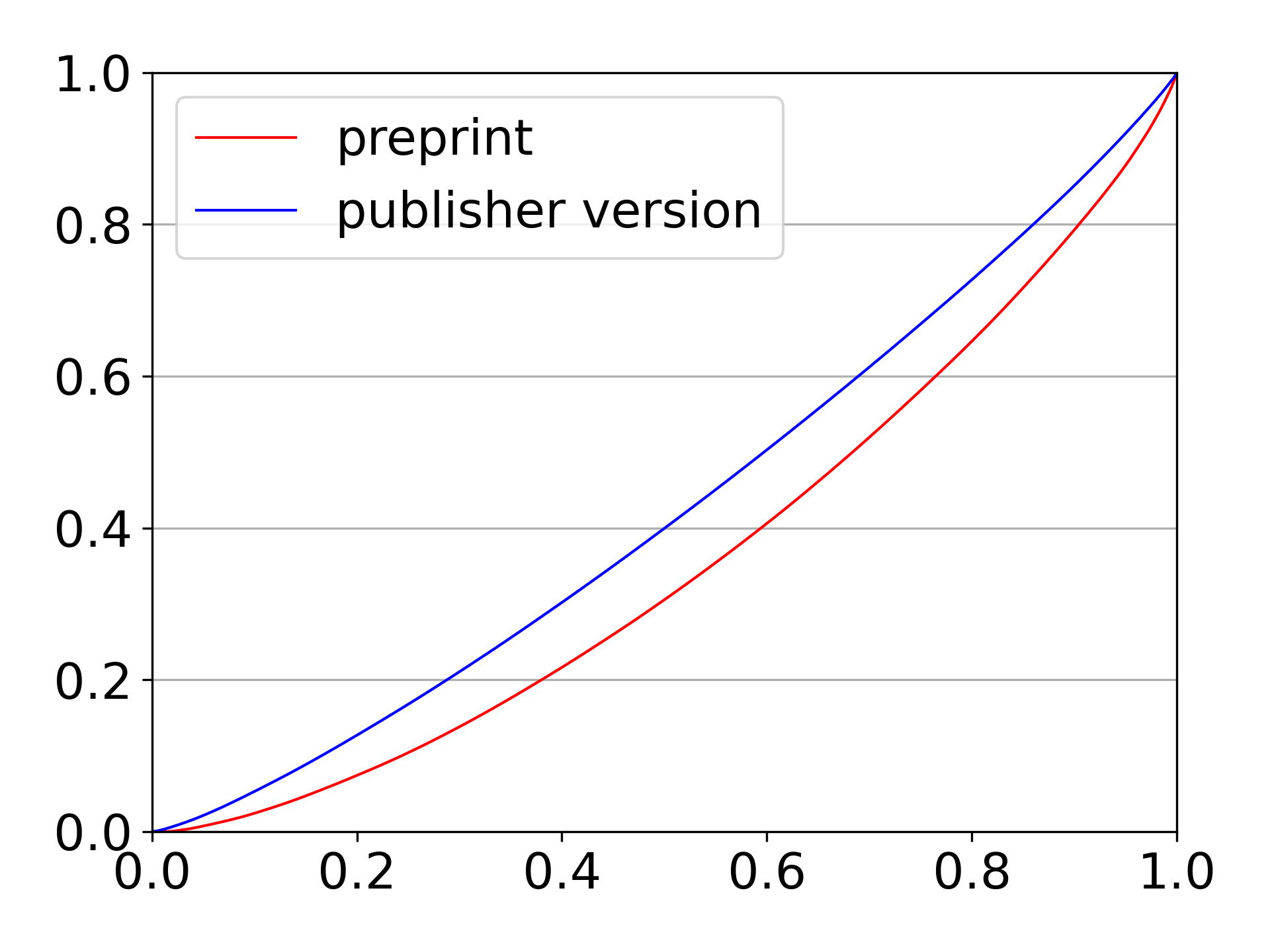}}%
	\subfigure[institution, last author]{%
		\includegraphics[clip, width=0.49\columnwidth]{figure/lorenz_first_institution_all_0-na_5_true_true_ln}}%
	\\
	\subfigure[institution, corres. author]{%
		\includegraphics[clip, width=0.49\columnwidth]{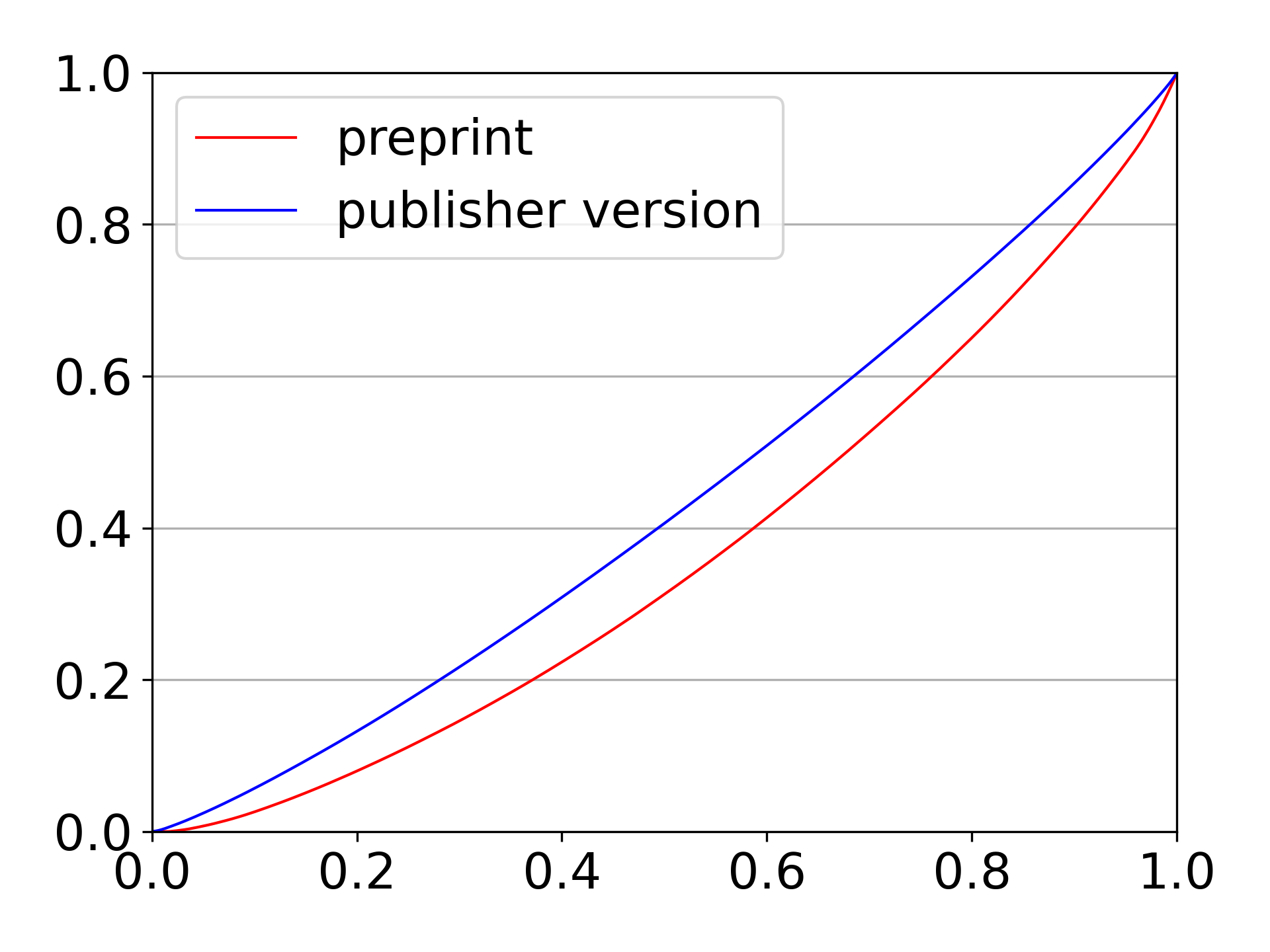}}%
	\subfigure[institution, all authors]{%
		\includegraphics[clip, width=0.49\columnwidth]{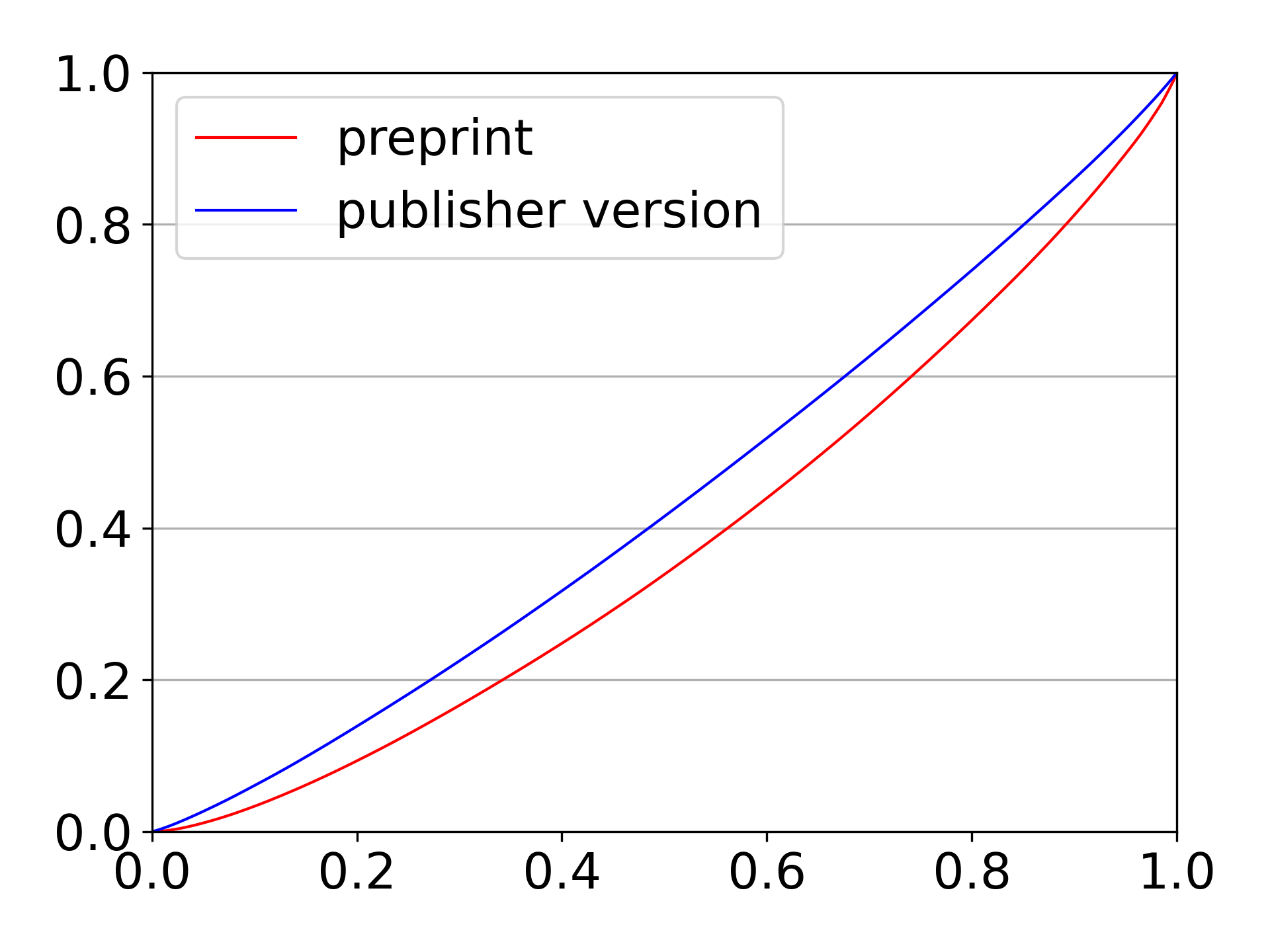}}%
	\\
	\subfigure[country, first author]{%
		\includegraphics[clip, width=0.49\columnwidth]{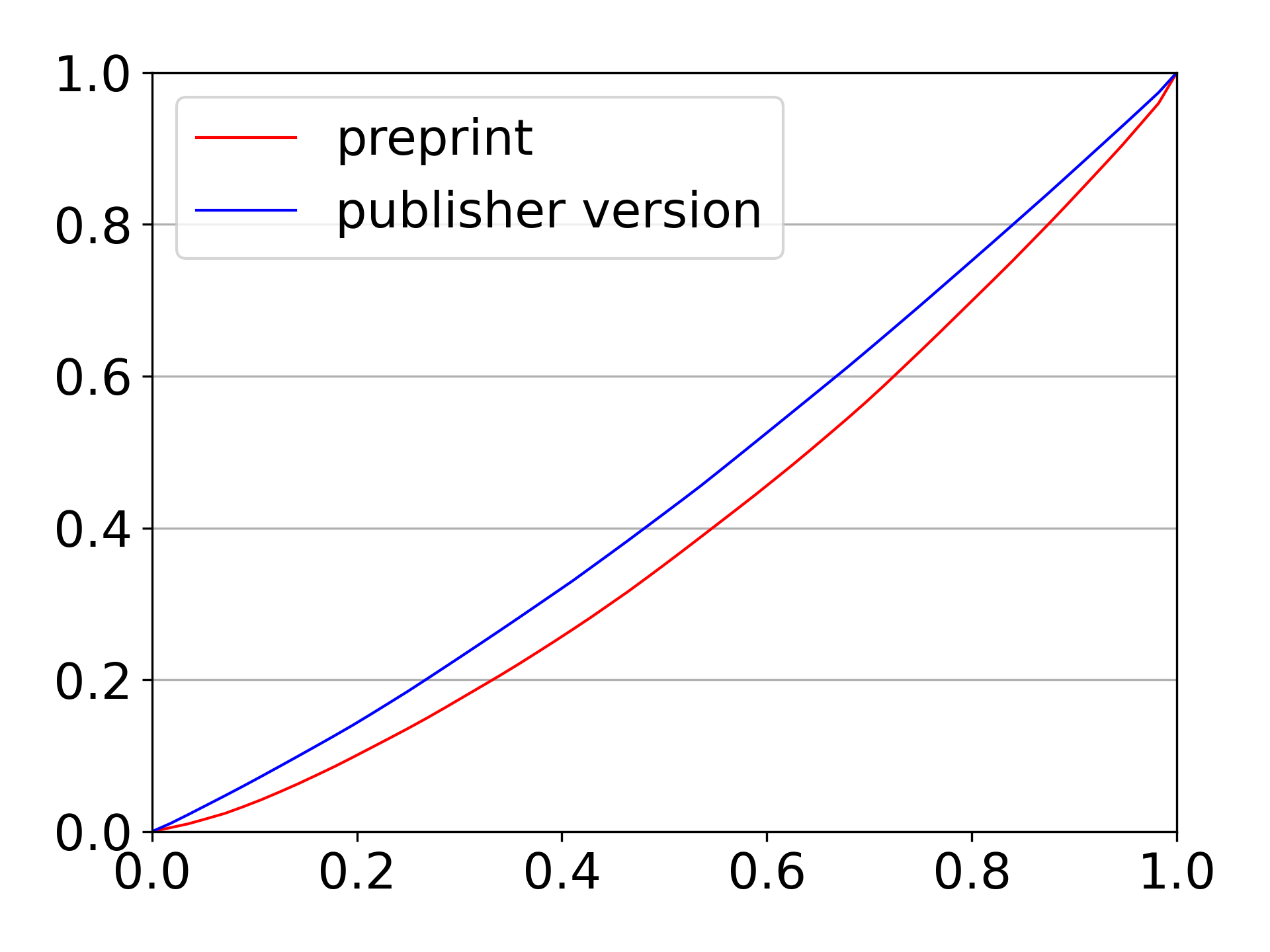}}%
	\subfigure[country, last author]{%
		\includegraphics[clip, width=0.49\columnwidth]{figure/lorenz_first_country_all_0-na_10_true_true_ln}}%
	\\
	\subfigure[country, corres. author]{%
		\includegraphics[clip, width=0.49\columnwidth]{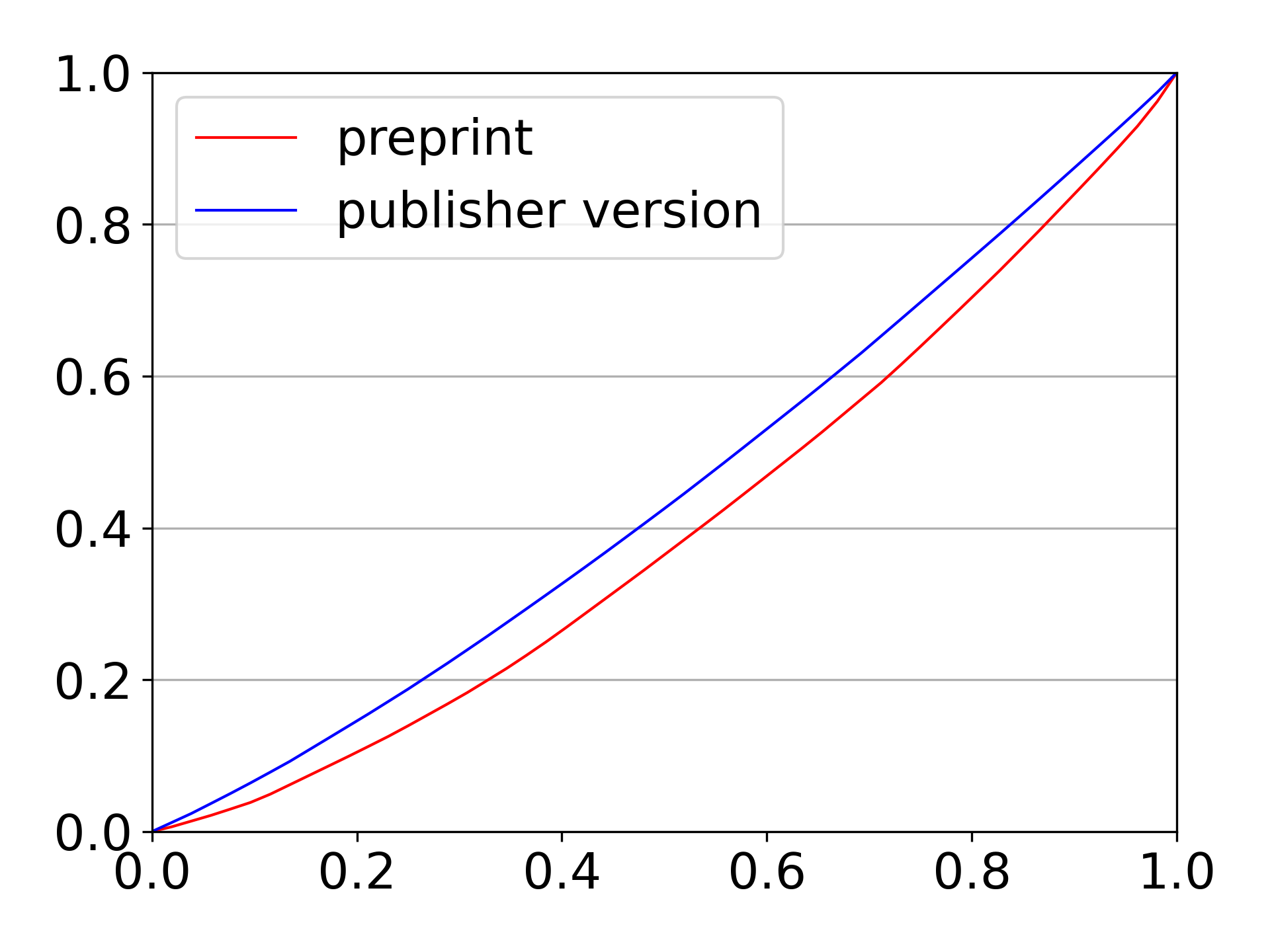}}%
	\subfigure[country, all authors]{%
		\includegraphics[clip, width=0.49\columnwidth]{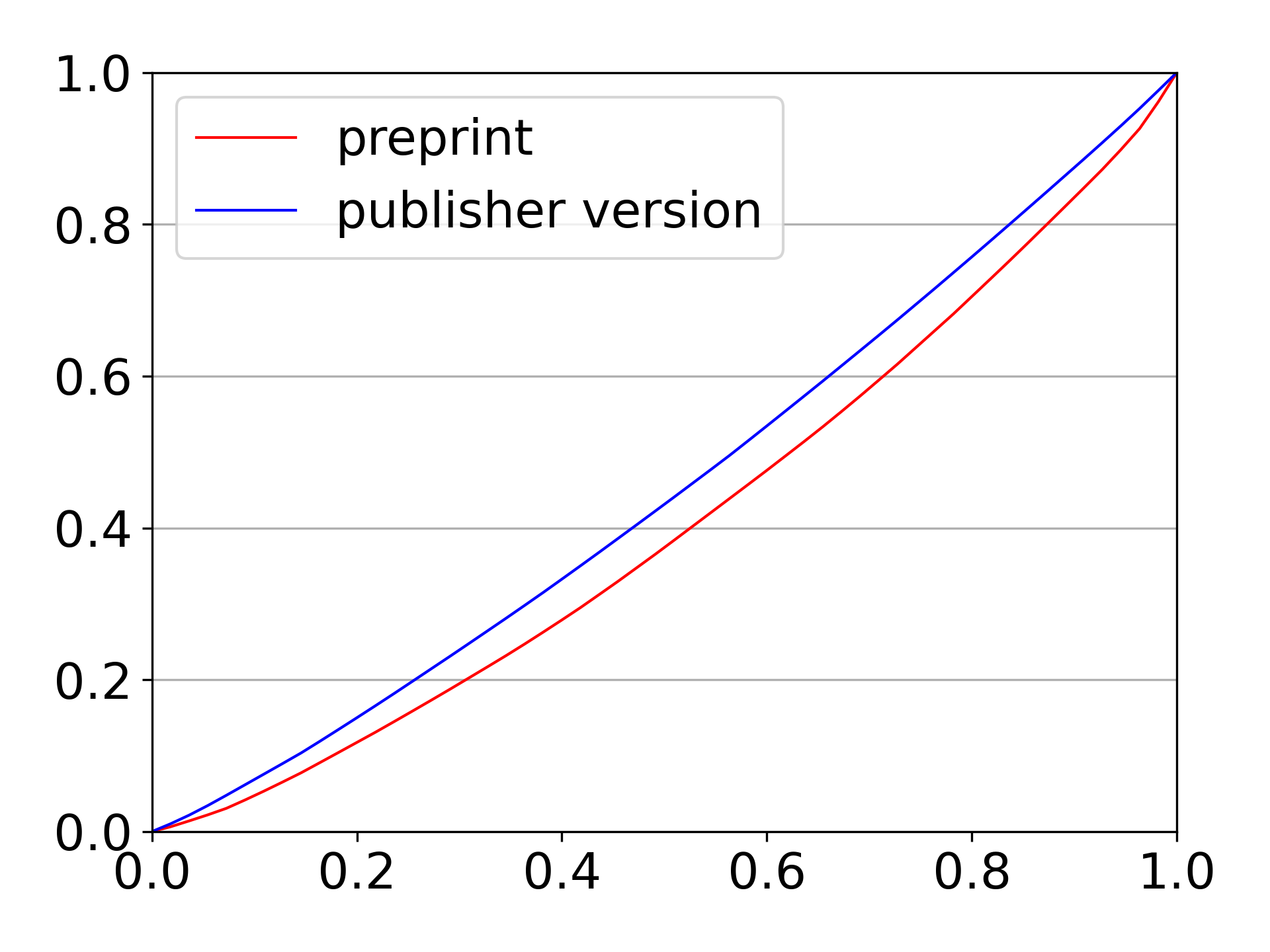}}%
	\caption{Lorenz curves of citations with different affiliation levels and target authors. }
	\label{fig:lorenz}
\end{figure}

\begin{table*}[tbp]
	\centering
	\caption{Gini coefficients of the number of citations with different affiliation levels and target authors. The third and fourth column present the Gini coefficients for preprints and publisher articles, respectively. The fifth column shows the absolute difference $\Delta$ between the Gini coefficients for preprints and publisher versions. The sixth and seventh column provide the number of articles (i\,.e., pairs of a preprint and its publisher version) and the number of affiliations that are involved in calculation of the Gini coefficients. }
	\begin{tabular}{l|l||r|r|r|r|r} \toprule
		\multicolumn{1}{c|}{affiliation level} & \multicolumn{1}{c||}{target author} & \multicolumn{1}{c|}{preprints} & \multicolumn{1}{c|}{pub. ver.} & \multicolumn{1}{c|}{$\Delta$} & \multicolumn{1}{c|}{\# of articles} & \multicolumn{1}{c}{\# of affiliations}
		\\ \midrule
		\multirow{4}{*}{institution} & first author & 0.28 & 0.15 & 0.14 & 24,023.67 & 835 \\ 
		 & last author & 0.28 & 0.14 & 0.14 & 23,876.76 & 823 \\
		 & corresp. author & 0.27 & 0.14 & 0.13 & 20.537.56 & 764 \\
		 & all authors & 0.23 & 0.12 & 0.11 & 24,035.25 & 824 \\ \hline
		 \multirow{4}{*}{country} & first author & 0.21 & 0.11 & 0.10 & 27,699.27 & 56 \\ 
		 & last author & 0.22 & 0.12 & 0.11 & 27,448.52 & 52 \\
		 & corresp. author & 0.19 & 0.10 & 0.09 & 23,997.16 & 52 \\
		 & all authors & 0.18 & 0.10 & 0.09 & 28,374.04 & 55 \\ 
		\toprule
	\end{tabular}
	\label{tb:gini}
\end{table*}

\section{Results}
\label{sec:results}

This section presents the results of our analysis. 
First, we show citations of preprints and their publisher versions. 
Thereafter, Section~\ref{sec:citation_inequality} presents citation inequality using the Lorenz curves and Gini coefficients. Sections~\ref{sec:influece_duration} and \ref{sec:influence_journals} verify the influence of duration between publications of preprint and publisher version and different journals on the results, respectively. 

\subsection{Citations of Preprints and Publisher Versions}

We first show in Figure~\ref{fig:time_articles_citations} how the number of citations of preprints and publisher versions evolve over time, starting from the publication month of the preprint.
In the upper graph of the figure, we observe an acceleration of the number of citations of preprints within the first 10 months following publications, and an approximate plateau between the months 10 and 24. 
On the other hand, the number of citations of publisher versions rises continuously over 24 months. 
The lower graph presents the number of preprints as well as publisher versions. 
We see that over half of the preprints have published their publisher versions within 8 months after their preprint publication. 

\subsection{Citation Inequality}
\label{sec:citation_inequality}

Figure~\ref{fig:lorenz} presents Lorenz curves with different affiliation levels (i\,.e., institution or country) and target authors (i\,.e., first author, last author, corresponding author, or all authors). Please note that we adopt fractional counting, where a co-authored article's citations are assigned fractionally to each of the co-authors' affiliations. In addition, if an author belongs to more than one affiliations, citations are apportioned to the affiliations. 
To eliminate the influence of affiliations with a small number of articles, we filter out institutions and countries that publish fewer than 5 and 10 articles, respectively, which are equivalent to approximately 30\% of preprints and their publisher versions. 
If we include filtered-out institutions and countries in the analysis, we observe even larger inequalities in both preprints and publisher versions and disparities between Lorenz curves for preprints and publisher versions. 

In Figure~\ref{fig:lorenz}, we consistently observe larger citation inequalities in preprints than in publisher versions. Larger disparities between Lorenz curves for preprints and publisher versions are shown on the institution level than on the country level.

Table~\ref{tb:gini} shows the Gini coefficients calculated based on the Lorenz curves shown in Figure~\ref{fig:lorenz}. 
The Gini coefficients are consistently higher for preprints than publisher versions. 
The coefficients for preprints are almost twice as those for publisher versions. 
Thus, there are larger citation inequalities in preprints than in publisher versions, and there could exist a larger citation bias in preprints. 
In addition, we consistently observe higher Gini coefficients on the institution level than on the country level. 
In other words, there is a greater imbalance in received citations across institutions than across countries.
The differences are smaller when we consider all authors, 
as the Gini coefficients for preprints get smaller.

We further investigate biased author affiliations, specifically countries, by examining differences in the ranks of the number of citations in preprints and publisher versions, considering all authors of the articles. 
Countries of authors in preprints seem to be more decisive than countries of authors in publisher versions. We rank countries in the order of the number of citations with respect to preprints and publisher versions. 
Countries that benefit from author affiliations (i.e., countries ranked higher in preprints) include the United States and the United Kingdom, which have the highest number of articles, but also developing countries such as the Kenya and Tanzania.\footnote{The full list of countries is as follows: United States, United Kingdom, Germany, Canada, Netherlands, Sweden, India, Israel, Brazil, Denmark, Norway, South Korea, Russia, South Africa, Chile, Greece, Iran, Ethiopia, Kenya, Colombia, Croatia, Uganda, Iceland, Tanzania.}
In contrast, countries ranked higher in publisher versions include Asian countries such as China, Japan, and Taiwan, and Latin American countries such as Mexico and Argentina.\footnote{The full list of countries is as follows: China, Australia, Switzerland, Japan, Italy, Finland, Singapore, Austria, New Zealand, Portugal, Czechia, Mexico, Argentina, Hungary, Ireland, Taiwan, Saudi Arabia, Turkey, Thailand, Malaysia, Estonia, Bangladesh, Nigeria, Slovenia, Luxembourg, Vietnam.} 

\subsection{Influence of Duration between Publications of Preprint and Publisher Version}
\label{sec:influece_duration}

There are possible factors that cause a bias in the Lorenz curves and Gini coefficients. 
One of them is the number of months from publication of a preprint to its publisher version. 
Peer-review is thought to improve the credibility of articles~\cite{soderberg2020credibility}, leading to increased citations. 
The distribution of the number of months from publication of a preprint to its publisher version varies greatly among journals and articles. 
If the number of articles by journal and the number of months from publication of a preprint to its publisher version are unbalanced among affiliations, the results shown in Figure~\ref{fig:lorenz} and Table~\ref{tb:gini} would be biased. 
Hence, we explore the Gini coefficients for preprints and publisher versions grouped by months from publication of a preprint to its publisher version. 
We consider institutions and countries that have published at least 3 and 5 articles at each month, respectively.

\begin{figure}[tb]
  \includegraphics[width=\columnwidth]{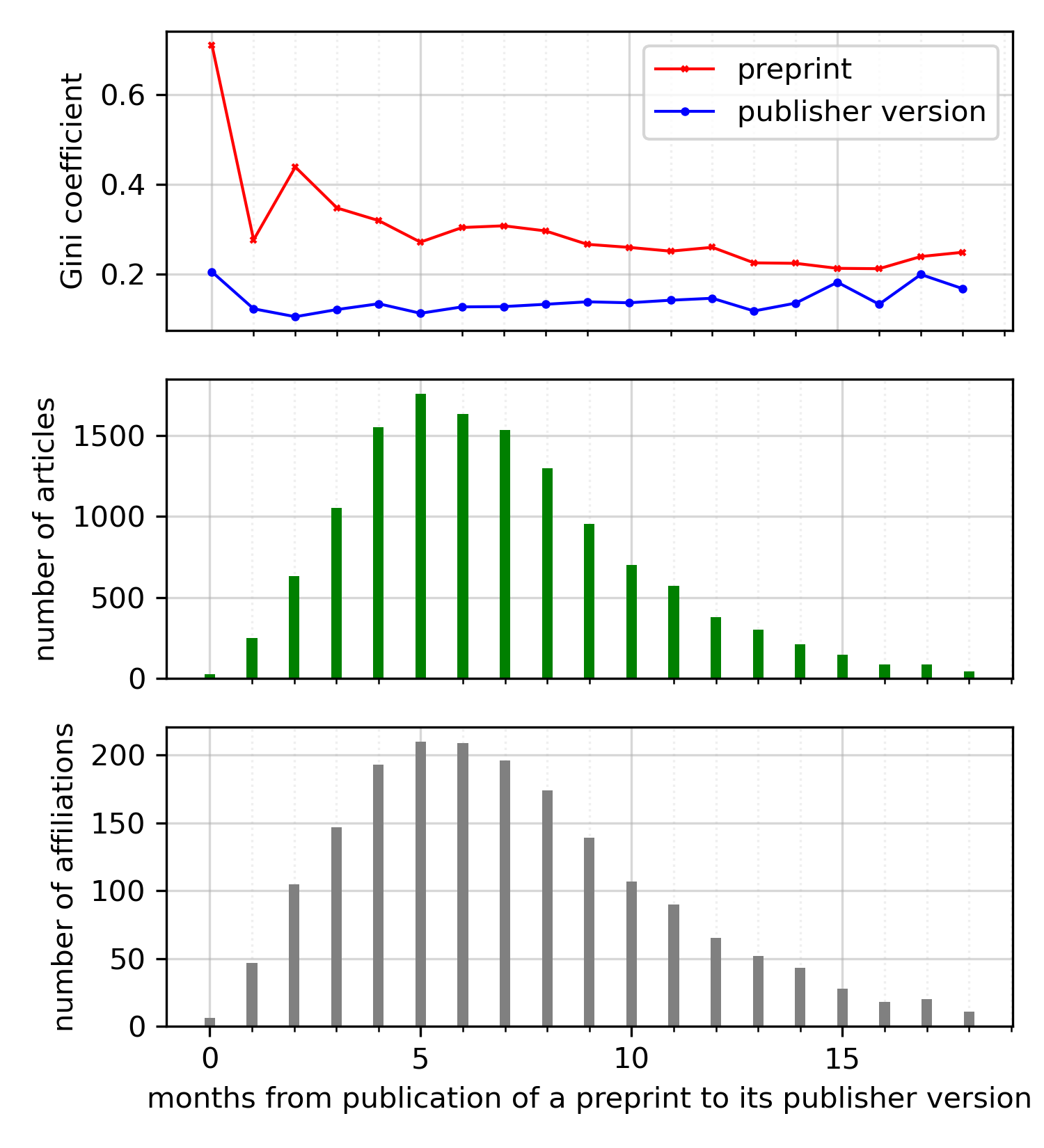}
  \caption{Gini coefficients of citations of preprints and publisher versions grouped by months from publication of a preprint to its publisher version at institution level.}
  \label{fig:gini_institution}
\end{figure}
\begin{figure}[tb]
  \includegraphics[width=\columnwidth]{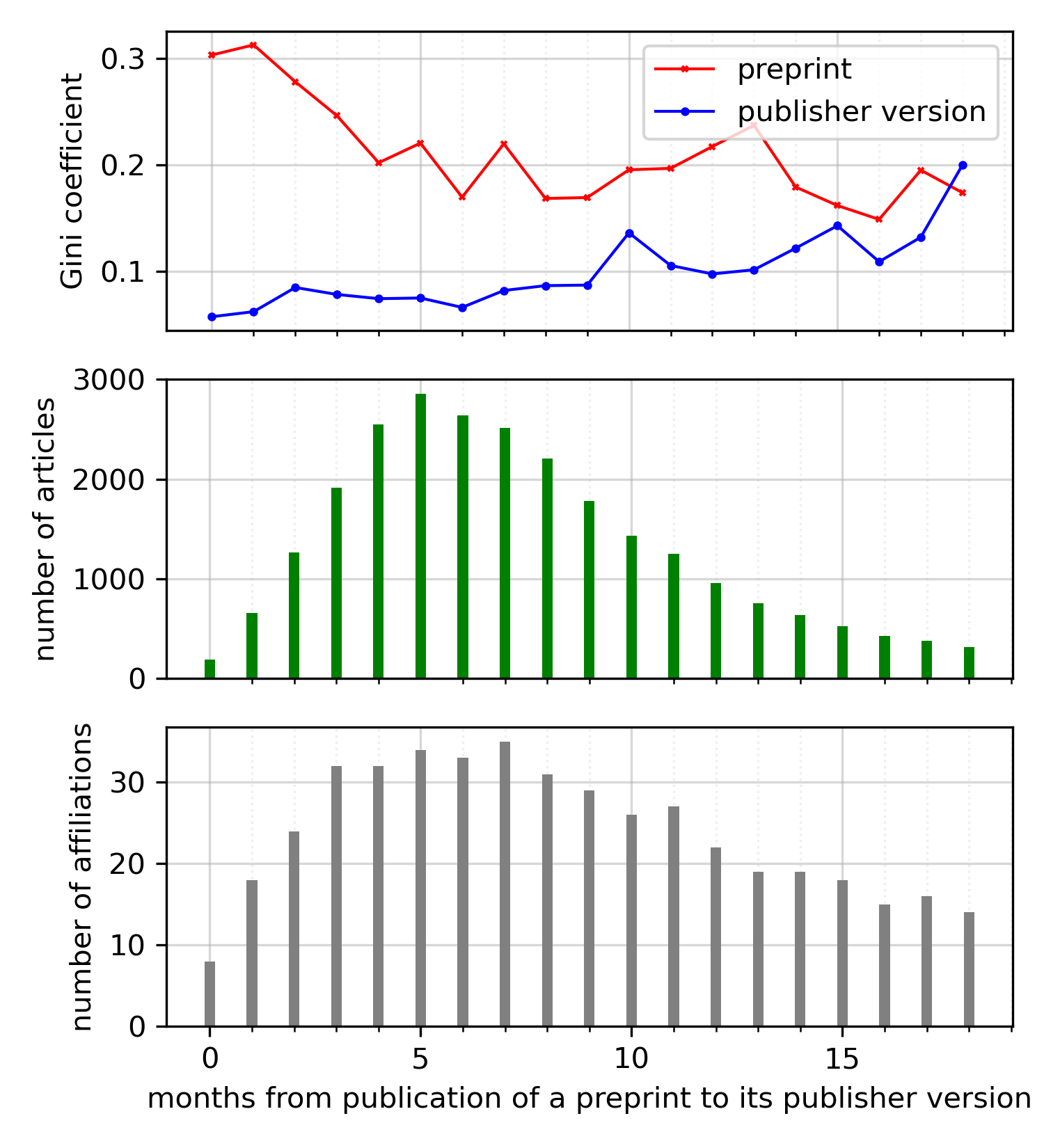}
  \caption{Gini coefficients of citations of preprints and publisher versions grouped by months from publication of a preprint to its publisher version at country level.}
  \label{fig:gini_country}
\end{figure}

Figures~\ref{fig:gini_institution} and \ref{fig:gini_country} show the Gini coefficients at institution and country levels, along with the number of articles and affiliations. 
In these figures, we set all authors as target authors. 
For instance, as can be seen in Figure~\ref{fig:gini_institution}, the Gini coefficient of citations of preprints that have been published 8 months after publication is 0.30. 
We observe larger Gini coefficients for preprints than those for publisher versions with one exception at the country level (e.g., articles that spend 18 months from publication of the preprint to its publisher version, as can be seen in Figure~\ref{fig:gini_country}). 
The differences between Gini coefficients get smaller as the duration between the publication of a preprint to its publisher version gets longer. This is because Gini coefficients for preprints get smaller while those for publisher versions become slightly larger. 
These tendencies are caused by the length of the observation period of citations. 
For example, if a publisher version has been published 5 months after preprint publication, the observation period of citations for the preprint and its publisher version is 4 months and 20 months, respectively. 
In a shorter observation period, the variance of the number of citations of the institution is larger. 
Even if the observation period is shorter, preprints have higher Gini coefficients than publisher versions. Hence, in our view, the influence of the length from publication of preprints compared to their publisher versions is limited. 

\subsection{Influence from Journals}
\label{sec:influence_journals}

\begin{table}[tbp]
	\centering
	\caption{Journals in which the publisher versions appear (descending according to the number of articles).} 
	\begin{tabular}{l|r} \toprule
journal & \# of articles \\ \midrule
PLOS ONE & 2,214 \\
Scientific Reports & 1,777 \\
eLife & 1,642 \\
Nature Communications & 1,493 \\
Proceedings of the National Academy of Sciences & 906 \\
PLoS Computational Biology & 789 \\
Bioinformatics & 773 \\
PLoS Genetics & 570 \\
Nucleic Acids Research & 500 \\
NeuroImage & 483 \\
		\toprule
	\end{tabular}
	\label{tb:journals}
\end{table}

\renewcommand{\arraystretch}{1.25} 
\begin{table*}[tb]
	\centering
	\caption{Gini coefficients of the number of citations in different journals and affiliation levels. The fifth and sixth column present the Gini coefficients for preprints and publisher articles. The seventh column shows the absolute difference $\Delta$ between the Gini coefficients for preprints and publisher versions. The eighth and ninth column provide the number of articles (i\,.e., pairs of a preprint and its publisher version) and the number of affiliations that are involved in calculation of the Gini coefficients. The tenth column presents mean average and standard deviation of the number of months between publication of preprint and publisher version. }
	\label{tb:gini_journals}
	\begin{tabularx}{\textwidth}{X|X|X|X||r|r|r|r|r|X} \toprule
journal type & journal name & JCR category & affiliation level & preprints & pub. ver. & $\Delta$ & \# articles & \# affiliations & \# months to publication of publisher version \\ \midrule
\multirow{4}{*}{mega-journal} & \multirow{2}{*}{PLOS ONE} &  \multirow{2}{*}{\shortstack[l]{Multidiscip.\\Science}} & institution & 0.83  & 0.31  & 0.52  & \multirow{2}{*}{1696.47}  & 2,107 & \multirow{2}{*}{6.92 (4.97)} \\
& & & country & 0.65  & 0.22  & 0.42  & & 107 & \\  \cline{2-10}
& \multirow{2}{*}{\shortstack[l]{Scientific\\Reports}} &  \multirow{2}{*}{\shortstack[l]{Multidiscip.\\Science}} & institution & 0.70  & 0.29  & 0.41  & \multirow{2}{*}{1,409.35}  & 1,623 & \multirow{2}{*}{8.92 (6.18)} \\
& & & country & 0.54  & 0.18  & 0.36  & & 89 & \\ \hline
\multirow{6}{*}{\shortstack[l]{disciplinary\\journal}} & \multirow{2}{*}{\shortstack[l]{Nucleic Acids\\Research}} &  \multirow{2}{*}{\shortstack[l]{Biochem. \&\\Mol. Biol.}} & institution & 0.61  & 0.26  & 0.34  & \multirow{2}{*}{388.46} & 539 & \multirow{2}{*}{7.96 (6.72)} \\
& & & country & 0.51  & 0.18  & 0.33  & & 47 & \\  \cline{2-10}
& \multirow{2}{*}{\shortstack[l]{Biophysical\\Journal}} & \multirow{2}{*}{Biophysics} & institution & 0.71  & 0.33  & 0.38  & \multirow{2}{*}{169.59} & 246 & \multirow{2}{*}{7.68 (4.93)} \\
& & & country & 0.60  & 0.20  & 0.40  & & 34 & \\  \cline{2-10}
& \multirow{2}{*}{\shortstack[l]{Nature\\Genetics}} & \multirow{2}{*}{\shortstack[l]{Genetics \&\\Heredity}} & institution & 0.30  & 0.13  & 0.17 & \multirow{2}{*}{150.82} & 678 & \multirow{2}{*}{10.67 (5.92)} \\
& & & country & 0.19  & 0.12  & 0.07  & & 53 & \\ \hline
\multirow{4}{*}{\shortstack[l]{prestigious\\journal}} & \multirow{2}{*}{Nature} & \multirow{2}{*}{\shortstack[l]{Multidiscip.\\Science}} & institution & 0.37  & 0.16  & 0.21 & \multirow{2}{*}{135.34} & 605 & \multirow{2}{*}{10.60 (7.24)} \\
& & & country & 0.27 & 0.10  & 0.18  & & 66 & \\ \cline{2-10}
& \multirow{2}{*}{Science} & \multirow{2}{*}{\shortstack[l]{Multidiscip.\\Science}} & institution & 0.33 & 0.14  & 0.19 & \multirow{2}{*}{110.19} & 455 & \multirow{2}{*}{7.79 (5.51)} \\
& & & country & 0.36 & 0.08  & 0.28 & & 51 & \\
		\toprule
	\end{tabularx}
\end{table*}

Another possible factor is the journal in which the publisher version appeared. 
The journal can be considered as a kind of indicator of the quality of an article. 
Thus, it has a considerable influence on the number of citations, thereby affecting the Lorenz curves and the magnitude of the Gini coefficients. 
However, if an article is cited solely based on its quality, it does not make a difference in the Lorenz curve and Gini coefficient of citations between preprints and the publisher versions with respect to affiliations.

Table~\ref{tb:journals} shows a large fraction of the 36,628 preprints have been published in mega-journals, a type of open access journal. 
What distinguishes mega-journals from other open access journals is that their peer-review process can solely focus on scientific trustworthiness~\cite{bjork2018evolution}, because they have no need to filter articles due to restricted numbers of slots in their publishing schedule~\cite{bjork2018evolution}.
PLOS ONE, Scientific Reports, eLife, and Nature Communications in Table~\ref{tb:journals} are considered as mega-journals~\cite{bjork2018evolution,mcgillivray2019relationship,wakeling2019motivations}. 

We investigate the citation inequality with respect to various journals of different types. 
We randomly select two mega-journals and three disciplinary journals from journals with at least 100 articles.
We also include the most prestigious multidisciplinary journals---i\,.e., Nature and Science---in the analysis. 
Table~\ref{tb:gini_journals} presents the Gini coefficients of citations in each of the selected journals. 
Again, we set all authors as target authors. 
As we do not filter affiliation by the number of articles, the Gini coefficients in Table~\ref{tb:gini_journals} are higher than those including all journals (see Table~\ref{tb:gini}). 
We consistently observe higher Gini coefficients for preprints than those for publisher versions in different journal types. 
Especially, the gap of citation inequality in mega-journals is large. 
The large gaps come from a large fraction of uncited preprints for mega-journals. 
For PLOS ONE, 82.20\% of preprints have been not cited. After the publication of the publisher versions, the percentage decreased to 16.71\%. 
This result aligns with Lou and He~\cite{lou2015does} and we observe even larger uncitedness in preprints than in publisher versions. 

\section{Discussion and Future Direction}
\label{sec:discussion}

In this paper, we explore citation bias in preprints associated with the author affiliations. The results of the analysis show larger citation inequalities in preprints than in publisher versions that indicate the author affiliations might influence the readership and the perception of preprints. The main difference between preprints and their respective publisher versions is the presence of peer-review process. Hence, the peer-review process mitigates citation bias.

However, as Tahamtan et al.~\cite{tahamtan2016factors} outlined, there are other factors that could influence on the number of citations. Other factors are obvious to be investigated in the future. 

In addition, we do not consider discrepancies between preprints and publisher versions. 
In other words, we assume that there are revisions at the same degree between all pairs of preprints and publisher versions.  
Klein et al.~\cite{klein2019comparing} investigated textual similarity of preprints and publisher versions using arXiv and bioRxiv, and reported that there are no significant difference between them. 
On the other hand, Oikonomidi et al.~\cite{oikonomidi2020changes} stated that the evidence components reported across preprints and publisher versions are not stable over time, focusing on COVID-19 research. 
If publisher versions authored by some institutions are revised and improved to a greater extent than those authored by the other institutions, the inequalities in the number of citations between affiliations could be explained by the amount of revisions. Hence, citation bias caused by author affiliations can be considered less than that shown in the previous section. 
We plan to include the influence from discrepancies between preprints and publisher versions in our analysis in the future. 

Specific preprint servers, such as bioRxiv, provide a comment function to users. The comment function enables quick feedback for the authors. Soderberg et al.~\cite{soderberg2020credibility} stated that 37\% of user study participants considered user comments as being extremely or moderately important. Furthermore, the comments might influence the users' judgments regarding whether they would read and cite the preprint. Therefore, we plan to investigate how the number of comments and the polarity of comments (i.e., positive or negative) influence the number of citations. 

\section{Conclusion}
\label{sec:conclusion}

In this paper, we examined the presence of citation bias in preprints and their publisher versions with respect to the articles' author affiliations. We observed larger citation inequalities in preprints than in publisher versions, indicating that author affiliations might influence the readership and the perception of preprints in general. The peer-review process mitigates this inequality. Ultimately, our study shows that authors need to be careful when citing works and that they should not be blinded by the author affiliation information. In addition, preprints increasingly attract attention in various cases, such as funding applications and recruitment. In these cases, funding agencies or referees would pay attention to citation-based metrics of preprints. As we observed even larger citation inequalities in preprints than in publisher versions, such institutions might want to be even more careful when using citations of preprints. 



\begin{acks}
This work was supported by JSPS KAKENHI Grant Number 20K20132. 
We also thank anonymous reviewers for their constructive feedback.  
\end{acks}

\bibliographystyle{ACM-Reference-Format}
\bibliography{ref}

\end{document}